\documentclass{ifacconf}

\usepackage{natbib}  

\usepackage{graphicx}
\usepackage{caption}

\usepackage{multicol}
\usepackage{amsmath,amssymb}
\usepackage{amsfonts}
\usepackage{textcomp}
\usepackage{psfrag}
\usepackage{multimedia}
\usepackage{fancybox}

\newcommand{\vect}[1]{\ensuremath{\boldsymbol{\mathrm{#1}}}}

\newtheorem{Proposition}{Proposition}

\newtheorem{Lemma}{Lemma}

\usepackage[usenames,dvipsnames]{color}
\definecolor{wheat}{rgb}{0.96,0.87,0.70}
\definecolor{mario}{rgb}{0.8,0.8,1}
\definecolor{seb}{rgb}{0.8,1,0.8}

\newcommand {\matr}[2]{\left[\begin{array}{#1}#2\end{array}\right]}

\newcounter{lastnote}

\begin{document} 
\begin{frontmatter}

\title{Reinforcement Learning for Mixed-Integer Problems Based on MPC} 

\author[First]{Sebastien Gros} 
\author[Second]{Mario Zanon} 

\address[First]{Norwegian University of Technology, NTNU}
\address[Second]{IMT School for Advanced Studies Lucca}




%
%


	\begin{abstract} Model Predictive Control has been recently proposed as policy approximation for Reinforcement Learning, offering a path towards safe and explainable Reinforcement Learning. This approach has been investigated for $Q$-learning and actor-critic methods, both in the context of nominal Economic MPC and Robust (N)MPC, showing very promising results. In that context, actor-critic methods seem to be the most reliable approach. Many applications include a mixture of continuous and integer inputs, for which the classical actor-critic methods need to be adapted. In this paper, we present a policy approximation based on mixed-integer MPC  schemes, and propose a computationally inexpensive technique to generate exploration in the mixed-integer input space that ensures a satisfaction of the constraints. We then propose a simple compatible advantage function approximation for the proposed policy, that allows one to build the gradient of the mixed-integer MPC-based policy.
	\end{abstract}
	

\end{frontmatter}

\section{Introduction}
Reinforcement Learning (RL) is a powerful tool for tackling stochastic processes without depending on a detailed model of the probability distributions underlying the state transitions. Indeed, most RL methods rely purely on observed data, and realizations of the stage cost assessing the system performance. RL methods seek to increase the closed-loop performance of the control policy deployed on the system as observations are collected. RL has drawn an increasingly large attention thanks to its accomplishments, such as, e.g., making it possible for robots to learn to walk or fly from experiments \citep{Wang2012b,Abbeel2007}. 

Most RL methods are based on learning the optimal control policy for the real system either directly, or indirectly. Indirect methods typically rely on learning a good approximation of the optimal action-value function underlying the system. The optimal policy is then indirectly obtained as the minimizer of the value-function approximation over the inputs. Direct RL methods, if based on the policy gradient, seek to adjust the parameters $\vect \theta$ of a given policy $\pi_{\vect\theta}$ such that it yields the best closed-loop performance when deployed on the real system. An attractive advantage of direct RL methods over indirect ones is that they are based on formal necessary conditions of optimality for the closed-loop performance of $\pi_{\vect\theta}$, and therefore guarantee - for a large enough data set - the (possibly local) assymptotic optimality of the parameters $\vect\theta$ \citep{Sutton1999, Silver2014}.

RL methods often rely on Deep Neural Networks (DNN) to carry the policy approximation $\pi_{\vect\theta}$. Unfortunately, control policies based on DNNs provide limited opportunities for formal verifications of the resulting policy, and for imposing hard constraints on the evolution of the state of the real system. The development of safe RL methods, which aims at tackling this issue, is currently an open field of research \citep{Garcia2015}. A novel approach towards providing formal safety certificates in the context of RL has been recently proposed in \citep{Gros2019,Gros2020a,Zanon2019b}, where the policy approximation is based on robust Model Predictive Control (MPC) schemes rather than unstructured function approximators like DNNs. The validity of this choice is discussed in details in \citep{Gros2019}. In \citep{Gros2020a}, methodologies to deploy direct RL techniques on MPC-based policy approximations are proposed. These methodologies are, however, restricted to continuous input spaces and therefore exclude integer decision variables, which are central in a number of applications.

In this paper, we propose an extension of the policy gradient techniques proposed in \citep{Gros2020a} to mixed-integer problems. A mixed-integer MPC is used as a policy approximation, and a policy gradient method adjusts the MPC parameters for closed-loop performance. We detail how the actor-critic method can be deployed in this specific context. In particular, we propose an asymptotically exact hybrid stochastic-deterministic policy approach allowing for computing the policy gradient at a lower computational complexity than a full stochastic approach. We then propose a hybrid compatible advantage-function approximator tailored to our formulation. We finally detail how the mixed-integer MPC can be differentiated at a low computational cost, using principles from parametric Nonlinear Programming, in order to implement the actor-critic method. The proposed method is illustrated on a simple example, allowing for an unambiguous presentation of the results.

The paper is structured as follows. Section \ref{sec:Background} provides background material on MDPs and RL. Section \ref{sec:MPCIntro} presents the construction of a mixed-integer stochastic policy using a mixed-integer MPC scheme to support the policy approximation. Section \ref{sec:SafeStochasticPolicy} details an actor-critic method tailored to the proposed formulation, and how the policy gradient can be estimated. A compatible advantage function approximation is proposed. Section \ref{sec:Computations} details how the mixed-integer MPC scheme can be efficiently differentiated. Section \ref{sec:Example} proposes an illustrative example, and Section \ref{sec:Conclusion} provides some discussions.

\section{Background} \label{sec:Background}
In the following, we will consider that the dynamics of the real system are described as a stochastic process on (possibly) continuous state-input spaces. 
We will furthermore consider (possibly) stochastic policies $\pi$, taking the form of probability densities:
\begin{align}
\label{eq:StochPolicyDef}
{\pi}\left[\vect{a}\,|\,\vect{s}\right]\,:\, \mathbb R^m \times \mathbb R^n\rightarrow \mathbb R_+,
\end{align}
denoting the probability density of selecting a given input $\vect a$ when the system is in a given state $\vect s$. Deterministic policies delivering $\vect a$ as a function of $\vect s$ will be labelled as:
\begin{align}
\label{eq:DetPolicyDef}
\vect\pi\left(\vect s\right) \,:\, \mathbb R^n \rightarrow \mathbb R^m. 
\end{align}
Any deterministic policy can be viewed as a stochastic one, having a Dirac function as a probability density (or unit function for discrete inputs), i.e., ${\pi}\left[\,\vect a\,|\,\vect{s}\,\right] = \delta\left(\vect a - \vect\pi\left(\vect s\right)\right).$

We consider a stage cost function $L(\vect s,\vect a) \in\mathbb{R}$ and a discount factor $\gamma \in [0,1]$, the performance of a policy $\pi$ is assessed via the total expected cost:
\begin{align}
\label{eq:Return}
J(\pi) = \mathbb{E}_{{ \pi}}\left[\left. \sum_{k=0}^\infty\, \gamma^k L(\vect s_k,\vect a_k)\,\right|\, \vect a_k \sim  \pi\left[\, .\,|\,\vect s_k\right]\, \right].
\end{align}
The optimal policy associated to the state transition, the stage cost $L$ and the discount factor $\gamma$ is deterministic and given by:
\begin{align}
\label{eq:OptimalPolicy}
\pi_\star =\mathrm{arg} \min_{ \pi}\, J(\pi).
\end{align}
The value function $V_{\vect\pi}$, action-value function $Q_{\vect\pi}$ and advantage functions $A_{\vect\pi}$ associated to a given policy $\pi$ are given by \citep{Bertsekas1995,Bertsekas1996,Bertsekas2007}:
\begin{subequations}
\label{eq:Bellman}
\begin{align}
V_{\vect\pi}\left(\vect s\right) &= \mathbb{E}\left[ L(\vect s,\vect a) + \gamma V_{\vect\pi}(\vect s_{+})\,|\, \vect s,\, \vect a\right], \label{eq:Bellman:Policy:0} \\
Q_{\vect\pi}\left(\vect s,\vect a\right) &= L(\vect s,\vect a) + \gamma \mathbb{E}\left[V_{\vect\pi}(\vect s_{+})\,|\, \vect s,\, \vect a\right],  \label{eq:MDP:Qfunction:Generic}\\
A_{\vect\pi}\left(\vect s,\vect a\right) &= Q_{\vect\pi}\left(\vect s,\vect a\right) - V_{\vect\pi}\left(\vect s\right), \label{eq:MDP:Afunction:Generic}
\end{align}
\end{subequations}
where the expected value in \eqref{eq:MDP:Qfunction:Generic} is taken over the state transition, and the one in \eqref{eq:Bellman:Policy:0} is taken over the state transitions and \eqref{eq:StochPolicyDef}. 
\subsection{Stochastic policy gradient} \label{sec:StochPolicyGradient}

In most cases, the optimal policy $\pi_\star$ cannot be computed, either because the system is not exactly known or because solving \eqref{eq:Bellman} is too expensive. 
It is then useful to consider approximations $\pi_{\vect\theta}$ of the optimal policy, parametrized by $\vect\theta$. The optimal parameters $\vect\theta_\star$ are then given by:
\begin{align}
\vect \theta_\star = \mathrm{arg}\min_{\vect\theta}\, J(\pi_{\vect\theta}). \label{eq:OptParameters}
\end{align}
The policy gradient $\nabla_{\vect{\theta}}\, J({\pi}_{\vect{\theta}}) $ associated to the stochastic policy $\pi_{\vect{\theta}}$ is then instrumental in finding $\vect \theta_\star$ by taking gradient steps in $\vect \theta$. The policy gradient can be obtained using various actor-critic methods~\citep{Sutton1998,Sutton1999}. In this paper, we will use the actor-critic formulation:
\begin{align}
\nabla_{\vect{\theta}}\, J({\pi}_{\vect{\theta}}) 
\label{eq:StochasticPiGradient}
&= \mathbb{E}_{{\vect{\pi}_{\vect{\theta}}}}\left[\nabla_{\vect{\theta}}\log {\pi}_{\vect{\theta}}\, A_{{\pi}_{\vect{\theta}}}\right], 
\end{align}
for stochastic policies, and the actor-critic formulation:
\begin{align}
\nabla_{\vect{\theta}}\, J({\vect\pi}_{\vect{\theta}}) 
\label{eq:DeterministicPiGradient}
&= \mathbb{E}_{{\vect{\pi}_{\vect{\theta}}}}\left[\nabla_{\vect{\theta}} {\vect\pi}_{\vect{\theta}}\, \nabla_{\vect a}A_{{\vect\pi}_{\vect{\theta}}}\right], 
\end{align}
for deterministic policies.

The value functions $V_{\vect\pi}$, $Q_{\vect\pi}$ and $A_{\vect\pi}$ associated to a given policy $\pi$ are typically evaluated via Temporal-Difference (TD) techniques \citep{Sutton1998}, and require that a certain amount of exploration is included in the deployment of the policy. For deterministic policies, the exploration can, e.g., be generated by including stochastic perturbations over the policy $\vect\pi_{\vect\theta}$, while stochastic policies generate exploration by construction. Note that is is fairly common in RL to define the stochastic policy $\pi_{\vect\theta}$ as an arbitrary density, e.g., the normal distribution, centered at a deterministic policy $\vect\pi_{\vect\theta}$. We shall observe here that the deterministic policy gradient \eqref{eq:DeterministicPiGradient} is not suited as such for integer inputs, as the gradients $\nabla_{\vect{\theta}} {\vect\pi}_{\vect{\theta}},\, \nabla_{\vect a}A_{{\vect\pi}_{\vect{\theta}}}$ do not exist on discrete input spaces. On continuous input spaces, the choice between the deterministic approach \eqref{eq:DeterministicPiGradient} or the stochastic approach \eqref{eq:StochasticPiGradient} is typically motivated by computational aspects.

\section{Mixed-integer Optimization-based policy} \label{sec:MPCIntro}

In this paper, we will consider parametrized deterministic policies $\vect{\pi}_{\vect\theta}\approx \vect\pi_\star$ based on parametric optimization problems. In particular, we will focus on optimization problems resulting from a nominal mixed-integer MPC formulation. The results proposed in this paper extend to robust MPC - enabling the construction of safe Reinforcement Learning methods - but this case is omitted in this paper for the sake of brevity.

 \subsection{Policy approximation based on mixed-integer MPC}
The mixed-integer MPC scheme reads as: 
\begin{subequations}
\label{eq:MPC:Policy:Constraints:MPC}
\begin{align}
\vect{u}^\star\left(\vect s,\vect{\theta}\right),&\, \vect{i}^\star\left(\vect{s},\vect{\theta}\right) =\nonumber\\ \mathrm{arg}\min_{\vect{u},\vect i}&\quad T(\vect x_{N},\vect\theta) +  \sum_{k=0}^{N-1} \ell(\vect x_{k},\vect u_{k},\vect i_{k},\vect\theta)\label{eq:Cost:MPC}\\
\mathrm{s.t.}&\quad \vect x_{k+1} = \vect F\left(\vect x_{k},\vect u_{k},\vect i_{k},\vect \theta\right),\,\,\, \vect x_{0} = \vect s, \label{eq:Dynamics:MPC}\\
&\quad  \vect{h}_k\left(\vect{x}_{k},\vect{u}_{k},\vect i_{k},\vect \theta\right) \leq 0, \quad k=0,...,N-1, \label{eq:PathConstraints:MPC} \\
&\quad  \vect{h}_N\left(\vect{x}_{N},\vect \theta\right) \leq 0, \label{eq:TerminalConstraints:MPC} \\
&\quad \vect i_k \in \left\{0,1\right\}^{m_{\mathrm i}}, \label{eq:Integer:MPC} 
\end{align}
\end{subequations}
where $\vect x_{k}\in\mathbb{R}^n$ are the predicted system trajectories, $\vect u_{k}\in \mathbb{R}^{m_{\mathrm c}}$ the planned continuous inputs and $\vect i_k \in \left\{0,1\right\}^{m_{\mathrm i}}$ the planned integer inputs. Without loss of generality, we consider binary integer inputs. Functions $\ell$, $T$ are the stage and terminal costs. Functions $\vect{h}_{0,\ldots,N-1}$ are the stage constraints and function $\vect{h}_N$ is the terminal constraint. 

For a given state $\vect s$ and parameters $\vect \theta$, the MPC scheme \eqref{eq:MPC:Policy:Constraints:MPC} delivers the continuous and integer input profiles
\begin{subequations}
\begin{align}
\vect{u}^\star\left(\vect s,\vect{\theta}\right) &= \left\{\vect u_{0}^\star\left(\vect s,\vect{\theta}\right),\ldots, \vect{u}_{N-1}^\star\left(\vect s,\vect{\theta}\right)\right\}, \\
\vect{i}^\star\left(\vect s,\vect{\theta}\right) &= \left\{\vect i_{0}^\star\left(\vect s,\vect{\theta}\right),\ldots, \vect{i}_{N-1}^\star\left(\vect s,\vect{\theta}\right)\right\},\end{align}
\end{subequations}
with $\vect{u}_{k}^\star\left(\vect s,\vect{\theta}\right)\in\mathbb{R}^{m_{\mathrm c}}$ and $\vect{i}_{k}^\star\left(\vect s,\vect{\theta}\right)\in\left\{0,1\right\}^{m_{\mathrm i}}$. 
The MPC scheme \eqref{eq:MPC:Policy:Constraints:MPC} generates a parametrized deterministic policy 
\begin{align}
\label{eq:Policy:Overall}
\vect{\pi}_{\vect\theta}\left(\vect s\right)= \left\{\vect{\pi}^\mathrm{c}_{\vect\theta}\left(\vect s\right),\, \vect{\pi}^\mathrm{i}_{\vect\theta}\left(\vect s\right)\right\},
\end{align}
where
\begin{subequations}
\begin{align}
\label{eq:Policy0}
\vect{\pi}^\mathrm{c}_{\vect\theta}\left(\vect s\right) &= \vect u_0^\star\left(\vect{s},\vect{\theta}\right) \,\in\,\mathbb{R}^{m_{\mathrm c}}, \\
\vect{\pi}^\mathrm{i}_{\vect\theta}\left(\vect s\right) &= \vect i_0^\star\left(\vect{s},\vect{\theta}\right) \,\in\,\left\{0,1\right\}^{m_{\mathrm i}},
\end{align}
\end{subequations}
are the first elements of the continuous and integer input sequences generated by \eqref{eq:MPC:Policy:Constraints:MPC}. In the following, it will be useful to consider the MPC scheme \eqref{eq:MPC:Policy:Constraints:MPC} as a generic parametric mixed-integer NLP:
\vspace{-0.1cm}
\begin{subequations}
\label{eq:Generic:NLP}
\begin{align}
\vect{u}^\star\left(\vect{s},\vect{\theta}\right),\, \vect{i}^\star\left(\vect{s},\vect{\theta}\right) = \mathrm{arg}\min_{\vect{u},\,\vect i}&\quad \Phi(\vect{x},\vect{u},\vect i,\vect{\theta})  \label{eq:Cost}\\
\mathrm{s.t.}&\quad \vect f\left(\vect x,\vect u,\vect i,\vect s,\vect{\theta}\right) = 0, \label{eq:DynamicConstraints}\\
&\quad \vect{h}\left(\vect{x},\vect{u},\vect i,\vect{\theta}\right) \leq 0, \label{eq:SafetyConstraints} \\
&\quad \vect i \in \left\{0,1\right\}^{m_{\mathrm i}\times N-1},
\end{align}
\end{subequations}
where function $\Phi$ gathers the stage and terminal cost functions from \eqref{eq:Cost:MPC}, function $\vect f$ gathers the dynamic constraints and initial conditions \eqref{eq:Dynamics:MPC}, and function $\vect h$ gathers the stage and terminal constraints \eqref{eq:PathConstraints:MPC}-\eqref{eq:TerminalConstraints:MPC}.

\section{actor-critic method} \label{sec:SafeStochasticPolicy}
In order to build actor-critic methods for \eqref{eq:Policy:Overall}, exploration is required \citep{Sutton1998}. When the input space is constrained and mixed-integer, the exploration becomes non-trivial to setup, as 1. it must retain the feasibility of the hard constraints \eqref{eq:PathConstraints:MPC}-\eqref{eq:TerminalConstraints:MPC} and 2. simple input disturbances are not possible for the integer part since they are locked on an integer grid. To address this issue, we will adopt a stochastic policy approach, well suited for the integer part, and consider its asymptotically equivalent deterministic counterpart on the continuous input space, well suited for computational efficiency.

\subsection{MPC-based exploration}\label{sec:PolicyFromMPC}
In order to generate exploration, we will build a stochastic policy \eqref{eq:StochPolicyDef} based on the deterministic policy \eqref{eq:Policy:Overall}
where $\vect a$ will gather the continuous inputs $\vect a^\mathrm{c}$ and integer inputs $\vect a^\mathrm{i}$ actually applied to the real system, i.e., $\vect a = \left\{\,\vect a^\mathrm{c},\,\vect a^\mathrm{i}\,\right\}$. We will build \eqref{eq:StochPolicyDef} such that it generates exploration that is respecting the constraints \eqref{eq:PathConstraints:MPC}-\eqref{eq:TerminalConstraints:MPC} with unitary probability. We propose to build \eqref{eq:StochPolicyDef} such that it becomes naturally separable between the integer and continuous part in the policy gradient computation. To that end, we consider a softmax approach to handle the integer part of the problem. More specifically, we consider the parametric mixed-integer NLP:
\begin{subequations}
\label{eq:Generic:NLP:InitialIntegerEmbedding}
\begin{align}
\Phi^\mathrm{i}(\vect s,\vect{\theta},\vect a^\mathrm{i})= \min_{\vect{u},\,\vect i}&\quad \Phi(\vect{x},\vect{u},\vect i,\vect{\theta}) \\
\mathrm{s.t.}&\quad \vect f\left(\vect x,\vect u,\vect i,\vect s,\vect{\theta}\right) = 0, \label{eq:DynamicConstraints:Stoch}\\
&\quad \vect{h}\left(\vect{x},\vect{u},\vect i,\vect{\theta}\right) \leq 0, \label{eq:SafetyConstraints:Stoch} \\
&\quad \vect i_0 = \vect a^\mathrm{i}, \label{eq:IntegerAssigned} \\
&\quad \vect i_{1,\ldots,N-1} \in \left\{0,1\right\}^{m_{\mathrm i}},
\end{align}
\end{subequations}
derived from \eqref{eq:Generic:NLP}, where the first integer input is assigned to $\vect a^\mathrm{i}$ via constraint \eqref{eq:IntegerAssigned}. We will consider that $\Phi^\mathrm{i}(\vect s,\vect{\theta},\vect a^\mathrm{i})$ takes infinite value when the selected integer input $\vect a^\mathrm{i}$ is infeasible. Let us label $\mathbb I(\vect s,\vect{\theta})$ the feasible set of $\vect a^\mathrm{i}$ for a given state $\vect s$ and MPC parameter $\vect\theta$, and $\tilde{\vect i}(\vect s,\vect{\theta},\vect a^\mathrm{i})$ the integer profile solution  of \eqref{eq:Generic:NLP:InitialIntegerEmbedding}. By construction $\tilde{\vect i}_0(\vect s,\vect{\theta},\vect a^\mathrm{i}) = \vect a^\mathrm{i}$ when $\vect a^\mathrm{i}\in\mathbb I(\vect s,\vect{\theta})$.
We then define the softmax stochastic integer policy distribution using
\begin{align}
\label{eq:Softmax}
\pi_{\vect\theta}^\mathrm{i}\left[\,\vect a^\mathrm{i}\, | \, \vect s\,\right]\,\propto\, e^{- \sigma_\mathrm{i}^{-1}\Phi^\star_\mathrm{i}(\vect s,\vect{\theta},\vect a^\mathrm{i})} \in\mathbb R_+,
\end{align}
where $\sigma_\mathrm{i} > 0$ is a parameter adjusting the variance of $\pi_{\vect\theta}^\mathrm{i}$.
In order to build the continuous part of the policy, we will consider the continuous part $\vect a^\mathrm{c}$ of the stochastic policy as conditioned on $\tilde{\vect i}$, and taking the form of a probability density:
\begin{align}
\label{eq:ContinuousStochPolicy}
 \pi^\mathrm{c}_{\vect\theta}\left[\,\vect a^\mathrm{c}\, \right| \, \tilde{\vect i}(\vect s,\vect{\theta},\vect a^\mathrm{i}),\,\vect s\,\left.\right]\in\mathbb R_+, 
\end{align}
which will be constructed from the parametric NLP:
\begin{subequations}
\label{eq:Generic:NLP:Embedding:Continuous}
\begin{align}
\tilde{\vect u}(\vect s,\vect{\theta},\vect i,\vect d)=\mathrm{arg} \min_{\vect{u}}&\quad \Phi(\vect{x},\vect{u},\vect i,\vect{\theta}) + \vect d^\top \vect u_0 \\
\mathrm{s.t.}&\quad \vect f\left(\vect x,\vect u,\vect i,\vect s,\vect{\theta}\right) = 0, \label{eq:DynamicConstraints:Det}\\
&\quad \vect{h}\left(\vect{x},\vect{u},\vect i,\vect{\theta}\right) \leq 0, \label{eq:SafetyConstraints:Det} 
\end{align}
\end{subequations}
derived from \eqref{eq:Generic:NLP}, but where the integer input profile is entirely assigned, and where $\vect d\in\mathrm R^{m_\mathrm{c}}$ is a random vector chosen as $\vect d\sim\mathcal{N}\left(0,\sigma_\mathrm{c}I \right)$. The random variable $\vect a_\mathrm{c}$ in \eqref{eq:ContinuousStochPolicy} will then be selected as:
\begin{align}
\vect a^\mathrm{c} = \tilde{\vect u}_0(\vect s,\vect{\theta},\tilde{\vect i}(\vect s,\vect{\theta},\vect a^\mathrm{i}),\vect d).
\end{align}
As previously observed in \citep{Gros2020a}, while $\pi^\mathrm{c}_{\vect\theta}$ is easy to sample, it is in general difficult to evaluate.

Because $\vect a^\mathrm{c}$ is conditioned on $\tilde{\vect i}$ and, therefore, $\vect a^\mathrm{i}$, the Kolmogorov definition of conditional probabilities entails that the overall stochastic policy \eqref{eq:StochPolicyDef} reads as the distribution:
\begin{align}
\label{eq:StochPolicyDef:Definition}
\pi_{\vect\theta}[\,\vect a\,|\,\vect s\,] =  \pi^\mathrm{c}_{\vect\theta}\left[\,\vect a^\mathrm{c}\, \left | \, \tilde{\vect i}(\vect s,\vect{\theta},\vect a^\mathrm{i}),\,\vect s\,\right.\right] \, \pi_{\vect\theta}^\mathrm{i}\left[\,\vect a^\mathrm{i}\, | \, \vect s\,\right].
\end{align}
We establish next a straightforward but useful result concerning the stochastic policy \eqref{eq:StochPolicyDef:Definition}. 
\begin{Lemma} \label{Lemma:Feasibiility}
The stochastic policy \eqref{eq:StochPolicyDef:Definition} generates input samples $\vect a$ that are feasible for the MPC scheme \eqref{eq:MPC:Policy:Constraints:MPC}.
\end{Lemma}
\begin{pf} 
	Because $\Phi^\mathrm{i}(\vect s,\vect{\theta},\vect a^\mathrm{i})=+\infty$ when $\vect a^\mathrm{i}\notin\mathbb I(\vect s,\vect{\theta})$, policy \eqref{eq:Softmax} selects feasible integer inputs $\vect a^\mathrm{i}$ with probability~1. Furthermore, NLP \eqref{eq:Generic:NLP:Embedding:Continuous} is feasible for all $\vect a^\mathrm{i}\in\mathbb I(\vect s,\vect{\theta})$ and all $\vect d$, such that its solution satisfies constraints \eqref{eq:DynamicConstraints}-\eqref{eq:SafetyConstraints}. As a result, the samples $\vect a^\mathrm{i}, \vect a^\mathrm{c}$ generated from \eqref{eq:StochPolicyDef:Definition} are guaranteed to be feasible. $\hfill\qed$
\end{pf}
The policy gradient associated to \eqref{eq:StochPolicyDef:Definition} can be computed using \eqref{eq:StochasticPiGradient}. Unfortunately, it has been observed that this approach is computationally expensive for continuous input spaces \citep{Gros2020a} when the policy is restricted by non-trivial constraints. Hence, we now turn to detailing how the policy gradient associated to policy \eqref{eq:StochPolicyDef:Definition} can be efficiently computed.
\subsection{Policy gradient}
Using policy \eqref{eq:StochPolicyDef:Definition}, the stochastic policy gradient is separable between the continuous and integer part and reads as:
\begin{align}
\label{eq:StochasticPolicyGradient}
\nabla_{\vect\theta} J\left(\pi_{\vect\theta}\right) &= \mathbb{E}_{\pi_{\vect\theta}}\left[\nabla_{\vect\theta}\log \pi_{\vect\theta} A_{\vect\pi_{\vect\theta}}\right] \\
&= \mathbb{E}_{\pi_{\vect\theta}}\left[\nabla_{\vect\theta}\log \pi^\mathrm{c}_{\vect\theta} A_{\vect\pi_{\vect\theta}}\right]  + \mathbb{E}_{\pi_{\vect\theta}}\left[\nabla_{\vect\theta}\log \pi^\mathrm{i}_{\vect\theta} A_{\vect\pi_{\vect\theta}}\right] , \nonumber 
\end{align}
where $A_{\vect\pi_{\vect\theta}}$ is the advantage function associated to the stochastic policy \eqref{eq:StochPolicyDef:Definition}. Using \eqref{eq:Softmax}, we then observe that the score function associated to the integer part of the policy is simply given by:
\begin{align}
\label{eq:ScoreDetail}
\nabla_{\vect\theta}\log \pi^\mathrm{i}_{\vect\theta}[\,\vect a^\mathrm{i}\,|\, \vect s\,]  =& -\frac{1}{\sigma_\mathrm{i}}\nabla_{\vect\theta}\Phi^\star_\mathrm{i}(\vect s,\vect{\theta},\vect a^\mathrm{i}) \\ &+ \frac{1}{\sigma_\mathrm{i}}\sum_{\vect i_0\in \mathbb I(\vect s,\vect{\theta})} \pi_{\vect\theta}^\mathrm{i}\left[\,\vect i_0\, | \, \vect s\,\right] \nabla_{\vect\theta}\Phi^\star_\mathrm{i}(\vect s,\vect{\theta},\vect i_0) \nonumber.
\end{align}

The computation of the policy gradient associated to the continuous part of the stochastic policy ought to be treated differently. Indeed, it has been observed in \citep{Gros2020a} that deterministic policy gradient methods are computationally more effective than stochastic ones for policy approximations on problems having continuous input and state spaces. Defining the deterministic policy for the continuous inputs $\vect a^\mathrm{c}$ as
\begin{align}
{\vect \pi}^\mathrm{c}_{\vect\theta}\left(\vect s,\vect i\right) = \tilde{\vect u}_0(\vect s,\vect{\theta},\vect i,0), \label{eq:StochToDet}
\end{align}
where $\tilde{\vect u}_0$ is the first element of the solution of \eqref{eq:Generic:NLP:Embedding:Continuous}, we consider the approximation \citep{Silver2014}
\begin{align}
\mathbb{E}_{\pi_{\vect\theta}}\left[\nabla_{\vect\theta}\log \pi^\mathrm{c}_{\vect\theta} A_{\vect\pi_{\vect\theta}}\right] \approx \mathbb{E}_{\pi_{\vect\theta}}\left[\nabla_{\vect\theta}{\vect \pi}^\mathrm{c}_{\vect\theta} \nabla_{\vect a^\mathrm{c}}A_{\vect\pi_{\vect\theta}}\right],
\end{align}
which is asymptotically exact for $\sigma_\mathrm{c} \rightarrow 0$ under some technical but fairly unrestrictive assumptions. We can then use the asymptotically exact hybrid policy gradient
\begin{align}
\label{eq:HybridPolicyGradient}
\widehat{\nabla_{\vect\theta} J\left(\pi_{\vect\theta}\right)} 
&= \mathbb{E}_{\pi_{\vect\theta}}\left[\nabla_{\vect\theta}\vect \pi^\mathrm{c}_{\vect\theta} \nabla_{\vect a^\mathrm{c}}A_{\vect\pi_{\vect\theta}}\right]  + \mathbb{E}_{\pi_{\vect\theta}}\left[\nabla_{\vect\theta}\log \pi^\mathrm{i}_{\vect\theta} A_{\vect\pi_{\vect\theta}}\right],
\end{align}
as a computationally effective policy gradient evaluation. The stochastic policy \eqref{eq:ContinuousStochPolicy} is then deployed on the system and generates exploration, while the deterministic policy \eqref{eq:StochToDet} is used to compute the policy gradient \eqref{eq:HybridPolicyGradient}. We propose next a compatible advantage function approximator for \eqref{eq:HybridPolicyGradient}, offering a systematic approximation of the advantage function $A_{\vect\pi_{\vect\theta}}$.
\subsection{Compatible advantage function approximation}
We note that the advantage function approximation
\begin{align}
\label{eq:StochasticApproximation}
\hat A_{\pi_{\vect\theta}} = \vect w^\top\nabla_{\vect\theta}\log \pi_{\vect\theta} = \vect w^\top\nabla_{\vect\theta}\log \pi^\mathrm{i}_{\vect\theta}+ \vect w^\top\nabla_{\vect\theta}\log \pi^\mathrm{c}_{\vect\theta},
\end{align}
is compatible by construction~\citep{Silver2014} for the stochastic policy gradient \eqref{eq:StochasticPolicyGradient}, in the sense that
\begin{align}
\nabla_{\vect\theta} J\left(\pi_{\vect\theta}\right) &= \mathbb{E}_{\pi_{\vect\theta}}\left[\nabla_{\vect\theta}\log \pi_{\vect\theta} \hat A_{\vect\pi_{\vect\theta}}\right]
\end{align}
holds if $\vect w$ is the solution of the Least-Squares problem
\begin{align}
\label{eq:LSForA}
\vect w = \mathrm{arg}\min_{\vect w}\,\frac{1}{2} \mathbb{E}_{\pi_{\vect\theta}}\left[\left( A_{\vect\pi_{\vect\theta}}-\hat A_{\vect\pi_{\vect\theta}}\right)^2\right].
\end{align}
Similarly, we seek a compatible advantage function approximation for the hybrid policy gradient \eqref{eq:HybridPolicyGradient}. We propose the hybrid advantage function approximation, inspired from \citep{Gros2020a}:
\begin{align}
\label{eq:HybridApproximation}
\hat A_{\pi_{\vect\theta}} = \vect w^\top\nabla_{\vect\theta}\log \pi^\mathrm{i}_{\vect\theta}+ \vect w^\top\textcolor{black}{\frac{1}{\sigma_\mathrm{c}}}\nabla_{\vect\theta} \pi^\mathrm{c}_{\vect\theta}M \left(\vect e- \vect c\right),
\end{align}
where we label $\vect e = \vect a^\mathrm{c} - {\vect\pi}^\mathrm{c}_{\vect\theta}$ the exploration performed on the continuous part of the input space $\mathbb{R}^{m_\mathrm{c}}$, and $M\in\mathbb R^{m_\mathrm{c}\times m_\mathrm{c}}$ is symmetric and $\vect c\in\mathbb R^{m_\mathrm{c}}$. We will show in the following proposition that for $M$ and $\vect c$ adequately chosen, the advantage function approximation \eqref{eq:HybridApproximation} is compatible with the policy gradient \eqref{eq:HybridPolicyGradient}.
\begin{Proposition} \label{Prop:Main}
The hybrid function approximation \eqref{eq:HybridApproximation} is asymptotically compatible, i.e.,
\begin{align}
\label{eq:MainResult}
\lim_{\sigma_\mathrm{c} \rightarrow 0}\widehat{\nabla_{\vect\theta} J\left(\pi_{\vect\theta}\right)} =&\,\,\lim_{\sigma_\mathrm{c} \rightarrow 0} \mathbb{E}_{\pi_{\vect\theta}}\left[\nabla_{\vect\theta}\vect \pi^\mathrm{c}_{\vect\theta} \nabla_{\vect a^\mathrm{c}}\hat A_{\vect\pi_{\vect\theta}}\right] \\& \qquad+ \mathbb{E}_{\pi_{\vect\theta}}\left[\nabla_{\vect\theta}\log \pi^\mathrm{i}_{\vect\theta} \hat A_{\vect\pi_{\vect\theta}}\right] \nonumber 
\end{align}
holds for $\vect w$ solution of \eqref{eq:LSForA} and for $M,\,\vect c$ chosen according to~\citep{Gros2020a}:
\begin{align}
\vect c &= \frac{\sigma_\mathrm{c}}{2}\sum_{i=1}^{n_{\vect a}} \frac{\partial^2  \tilde{\vect u}_0}{\partial \vect d_i^2}, \quad
M=  \left(\frac{\partial  \tilde{\vect u}_0}{\partial \vect d}\frac{\partial \tilde{\vect u}_0}{\partial \vect d}^\top\right),  \label{eq:CovEstimator}
\end{align}
evaluated at the solution of \eqref{eq:Generic:NLP:Embedding:Continuous} for $\vect d = 0$, where \eqref{eq:Generic:NLP:Embedding:Continuous} satisfies the regularity assumptions of \citep[Proposition 1]{Gros2020a}. These assumptions are technical but fairly unrestrictive, see \citep{Gros2020a} for a complete discussion.
\end{Proposition}
The proof delivered below is a sketch that follows the lines of the proof of Proposition 1 in \cite{Gros2020a}.
\begin{pf}  
	We observe that the solution $\vect w$ of \eqref{eq:LSForA} using \eqref{eq:HybridApproximation} is given by:
\begin{align}
&\mathbb{E}_{\pi_{\vect\theta}} \hspace{-3pt} \left[\hspace{-1pt}\left( \hspace{-1pt} \nabla_{\vect\theta}\log \pi^\mathrm{i}_{\vect\theta}+\textcolor{black}{\frac{1}{\sigma_\mathrm{c}}}\nabla_{\vect\theta} \pi^\mathrm{c}_{\vect\theta}M \left(\vect e- \vect c\right)\hspace{-1pt} \right) \hspace{-3pt} \left( \hspace{-1pt} A_{\vect\pi_{\vect\theta}}-\hat A_{\vect\pi_{\vect\theta}} \hspace{-1pt}\right)\hspace{-1pt}\right] 
= 0. \label{eq:LSStationarity}
\end{align}
Using a Taylor expansion of $A_{\vect\pi_{\vect\theta}}$ at $\vect e=0$, as proposed in \citep[Proposition 1]{Gros2020a}, we observe that \eqref{eq:LSStationarity} becomes:
\begin{align}
\label{eq:AExpansion}
&\mathbb{E}_{\pi_{\vect\theta}}\left[ \nabla_{\vect\theta}\log \pi^\mathrm{i}_{\vect\theta} \left( A_{\vect\pi_{\vect\theta}}-\hat A_{\vect\pi_{\vect\theta}}\right)\right] +  \textcolor{black}{\frac{1}{\sigma_\mathrm{c}}}\mathbb{E}_{\pi_{\vect\theta}}\left[\nabla_{\vect\theta} \pi^\mathrm{c}_{\vect\theta}M \left(\vect e- \vect c\right) \xi\right] \nonumber\\ &+\mathbb{E}_{\pi_{\vect\theta}}\left[\nabla_{\vect\theta} \pi^\mathrm{c}_{\vect\theta}M \textcolor{black}{\frac{\vect e- \vect c}{\sigma_\mathrm{c}}} \left( A_{\vect\pi_{\vect\theta}}-\hat A_{\vect\pi_{\vect\theta}}\right)\right] +\\&+ \mathbb{E}_{\pi_{\vect\theta}}\left[\nabla_{\vect\theta} \pi^\mathrm{c}_{\vect\theta}M \textcolor{black}{\frac{\left(\vect e- \vect c\right)\vect e^\top}{\sigma_\mathrm{c}}}\left( \nabla_{\vect a^\mathrm{c}}A_{\vect\pi_{\vect\theta}}-\nabla_{\vect a^\mathrm{c}}\hat A_{\vect\pi_{\vect\theta}} \right)\right] =0, \nonumber
\end{align}
where $\xi$ is the second-order remainder of the Taylor expansion of $A_{\vect\pi_{\vect\theta}}$. Unlike \eqref{eq:LSStationarity}, all terms in \eqref{eq:AExpansion} are evaluated at $\vect s,\,\vect a^\mathrm c = \vect\pi^\mathrm c \left(\vect s\right)$. Following a similar argumentation as in \citep[Proposition 1]{Gros2020a}, we obtain 
\begin{subequations}
\begin{align}
\lim_{\sigma_\mathrm{c}\rightarrow 0} &\mathbb{E}_{\pi_{\vect\theta}}\left[\textcolor{black}{\frac{1}{\sigma_\mathrm{c}}}\nabla_{\vect\theta} \pi^\mathrm{c}_{\vect\theta}M \left(\vect e- \vect c\right)\vect e^\top\left( \nabla_{\vect a^\mathrm{c}}A_{\vect\pi_{\vect\theta}}-\nabla_{\vect a^\mathrm{c}}\hat A_{\vect\pi_{\vect\theta}} \right)\right] \nonumber \\
&=\lim_{\sigma_\mathrm{c}\rightarrow 0}  \mathbb{E}_{\pi_{\vect\theta}}\left[\nabla_{\vect\theta} \pi^\mathrm{c}_{\vect\theta}\left( \nabla_{\vect a^\mathrm{c}}A_{\vect\pi_{\vect\theta}}-\nabla_{\vect a^\mathrm{c}}\hat A_{\vect\pi_{\vect\theta}} \right)\right],  \label{eq:ShittyTerm0} \\
\lim_{\sigma_\mathrm{c}\rightarrow 0} &\mathbb{E}_{\pi_{\vect\theta}}\left[\textcolor{black}{\frac{1}{\sigma_\mathrm{c}}}\nabla_{\vect\theta} \pi^\mathrm{c}_{\vect\theta}M \left(\vect e- \vect c\right) \xi\right] = 0, \label{eq:ShittyTerm}\\
\lim_{\sigma_\mathrm{c}\rightarrow 0} &\mathbb{E}_{\pi_{\vect\theta}}\left[\nabla_{\vect\theta} \pi^\mathrm{c}_{\vect\theta}M \textcolor{black}{\frac{\vect e- \vect c}{\sigma_\mathrm{c}}} \left( A_{\vect\pi_{\vect\theta}}-\hat A_{\vect\pi_{\vect\theta}}\right)_{\vect e = 0}\right] = 0. \label{eq:ShittyTerm2}
\end{align}
\end{subequations}
Equality \eqref{eq:ShittyTerm} holds from the Delta method, while equalities \eqref{eq:ShittyTerm0}, \eqref{eq:ShittyTerm2} hold because
\begin{align}
\lim_{\sigma_\mathrm{c}\rightarrow 0}\mathbb{E}_{\pi_{\vect\theta}}\left[\frac{1}{\sigma_\mathrm{c}}M \left(\vect e- \vect c\right)\vect e^\top\right] = I, \\\lim_{\sigma_\mathrm{c}\rightarrow 0}\mathbb{E}_{\pi_{\vect\theta}}\left[M \textcolor{black}{\frac{\vect e- \vect c}{\sigma_\mathrm{c}}}\right] = 0,
\end{align}
result from \eqref{eq:CovEstimator}, see \citep{Gros2020a}. Hence
\begin{align}
\label{eq:FinalResult}
&\lim_{\sigma_\mathrm{c} \rightarrow 0}\mathbb{E}_{\pi_{\vect\theta}}\left[ \nabla_{\vect\theta}\log \pi^\mathrm{i}_{\vect\theta} \left( A_{\vect\pi_{\vect\theta}}-\hat A_{\vect\pi_{\vect\theta}}\right)\right]  \\ &\hspace{1.5cm} +\mathbb{E}_{\pi_{\vect\theta}}\left[\nabla_{\vect\theta} \pi^\mathrm{c}_{\vect\theta}\left( \nabla_{\vect a^\mathrm{c}}A_{\vect\pi_{\vect\theta}}-\nabla_{\vect a^\mathrm{c}}\hat A_{\vect\pi_{\vect\theta}} \right)\right] =0. \nonumber
\end{align}
Using \eqref{eq:HybridPolicyGradient}, \eqref{eq:MainResult} holds from \eqref{eq:FinalResult}.
$\hfill\qed$
\end{pf}

\section{NLP sensitivities} \label{sec:Computations}
In order to deploy the policy gradient techniques described above, one needs to compute the sensitivities $\nabla_{\vect\theta}\vect \pi^\mathrm{c}_{\vect\theta}$ and $\nabla_{\vect\theta}\log \pi^\mathrm{i}_{\vect\theta}$. 
Computing the score function \eqref{eq:ScoreDetail} requires computing the sensitivity of the cost function $\Phi^\star_\mathrm{i}$ of the NLP \eqref{eq:Generic:NLP:InitialIntegerEmbedding}. This sensitivity exists almost everywhere and is given by:
\begin{align}
 \nabla_{\vect\theta}\Phi^\star_\mathrm{i}(\vect s,\vect{\theta},\vect a^\mathrm{i}) =  \nabla_{\vect\theta}\mathcal {L}(\vect y,\vect\lambda, \vect\mu,\vect s,\vect{\theta},\vect a^\mathrm{i},\vect d),
\end{align}
where $\vect y$ is the primal solution of the NLP \eqref{eq:Generic:NLP:InitialIntegerEmbedding}, gathering the continuous inputs and states of the NLP, and $\vect\lambda, \vect\mu$ the dual variables associated to constraints \eqref{eq:DynamicConstraints}-\eqref{eq:SafetyConstraints}, respectively, and
$\mathcal {L} = \Phi + \vect d^\top \vect u_0 + \vect\lambda^\top \vect f + \vect\mu^\top \vect h$
is the Lagrange function associated to \eqref{eq:Generic:NLP:InitialIntegerEmbedding}. The computation of $\nabla_{\vect\theta} \vect \pi^\mathrm{c}_{\vect\theta}$ is more involved. Consider:
\begin{align}
\vect r = \matr{c}{\nabla_{\vect y}\mathcal {L}(\vect z,\vect s,\vect{\theta},\vect a^\mathrm{i},\vect d) \\
\vect f\left(\vect w,\vect s,\vect\theta\right)\\
\mathrm{diag}(\vect\mu)\vect h\left(\vect w,\vect\theta\right)+\tau
} = 0,
\end{align}
i.e., the primal-dual interior-point KKT conditions associated to  \eqref{eq:Generic:NLP:InitialIntegerEmbedding} for a barrier parameter $\tau>0$, and $\vect z$ gathering the primal-dual variables of the NLP \eqref{eq:Generic:NLP:InitialIntegerEmbedding}, i.e., 
 $\vect z = \left\{\vect y,\vect\lambda, \vect\mu\right\}$.
Then, if the solution of the NLP \eqref{eq:Generic:NLP:InitialIntegerEmbedding} satisfies LICQ and SOSC \citep{Nocedal2006}, the sensitivity of the solution of the NLP \eqref{eq:Generic:NLP:InitialIntegerEmbedding} exists almost everywhere and can be computed via the Implicit Function Theorem, providing
\begin{align}
\frac{\partial \vect z}{\partial \vect\theta} =  - \frac{\partial \vect r}{\partial \vect z}^{-1} \frac{\partial \vect r}{\partial \vect\theta}, 
\end{align}
see \citep{Buskens2001}. Using \eqref{eq:StochToDet}, the sensitivity $\nabla_{\vect\theta}\vect \pi^\mathrm{c}_{\vect\theta}$ then read as 
\begin{align}
\nabla_{\vect\theta} \vect \pi^\mathrm{c}_{\vect\theta} = \nabla_{\vect\theta}\tilde{\vect u}_0,
\end{align}
where $\nabla_{\vect\theta}\tilde{\vect u}_0$ is extracted from $\frac{\partial \vect z}{\partial \vect\theta}$.

\section{Simulated example} \label{sec:Example}
For the sake of brevity and in order to present results that are easy to interpret and verify, we propose to use a very low dimensional example, allowing us to bypass the evaluation of the action-value function via Temporal-Difference techniques, and isolate the discussions of this paper from questions regarding TD methods. We consider the linear, scalar dynamics:
\begin{align}
s_{k+1} = s_k + a^{\mathrm c}_k \, i_k + n_k
\end{align}
where $s_k, a^{\mathrm c}_k \in\mathbb{R}$, $i_k\in\left\{0,\,1\right\}$ and $n_k$ is uniformly distributed in $[0,\,0.05]$. We consider the baseline stage cost: 
\begin{align}
\label{eq:Example:BaselineCost}
L\left(s,\vect a\right) =& \frac{1}{2}(s-s_\mathrm{ref})^2 + \frac{1}{2}(a^{\mathrm c}-a^{\mathrm c}_\mathrm{ref})^2 + w\, i \\&\qquad 
+ c\,  \max\left(|s|-0.2,0\right), \nonumber
\end{align}
as the reference performance, where $w,\, c\in\mathbb{R}_+$ are scalar weight and $s_\mathrm{ref},\, a_\mathrm{ref}$ are references for the state and continuous input. The MPC model is deterministic, given by:
\begin{align}
x_{k+1} = x_k + u_k\, i_k + b
\end{align}
where $b\in\mathbb R$ is constant, but subject to adaptation via RL. The baseline cost imposes a high penalty for $s\notin [-0.2,0.2]$, and constitutes an exact relaxation of the constraint $-0.2\leq s\leq 0.2$, see \citep{Gros2019}. The MPC stage cost $\ell$ has the form \eqref{eq:Example:BaselineCost}. The MPC parameters $s_\mathrm{ref},\, a^{\mathrm c}_\mathrm{ref}$, $c$ and $b$ are subject to adaptation via RL.

The policy gradient \eqref{eq:MainResult} was implemented, where the advantage function estimation was computed from \eqref{eq:LSForA}, using the approximator $\hat A_{\pi_{\vect\theta}}$ from \eqref{eq:HybridApproximation}. The true advantage function $A_{\vect\pi_{\vect\theta}}$ was evaluated via classic policy evaluation \citep{Sutton1998} in order to deliver unambiguous results. On more complex examples \eqref{eq:LSForA} would be evaluated via Temporal-Difference techniques. The evaluations of \eqref{eq:MainResult} and  \eqref{eq:LSForA} were performed in a batch fashion, using 30 batches of 50 time steps each, all starting from the deterministic initial condition $\vect s_0=0$. The MPC scheme had a horizon of $N=10$ time samples, and a terminal cost based on the Riccati matrix of the control problem with $i = 1$. \textcolor{black}{A discount factor of $\gamma = 0.95$} was adopted. The step-size selected for adapting the parameters from the policy gradient was $\alpha = 2\cdot 10^{-3}$. The exploration parameters were chosen as $\sigma_\mathrm{i} = 2\cdot 10^{-2}$, $\sigma_\mathrm{c} = 10^{-2}$.

\begin{figure}
\center
\psfrag{x}[Bc][Bc][1.1]{$s$}
\psfrag{u}[Bc][Bc][1.1]{$a^{\mathrm c}$}
\psfrag{i}[Bc][Bc][1.1]{$i$}
\psfrag{time}[Bc][Bc][0.7]{time}
\includegraphics[width=0.46\textwidth,clip]{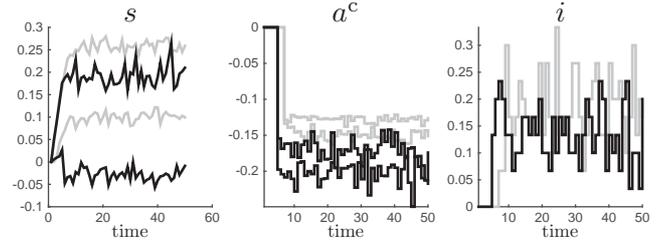}
\caption{Closed-loop trajectories before (light grey) and after (black) the learning. The left graph shows the extreme values of the state trajectories, the middle graph shows the extreme values of the continuous input $a^{\mathrm c}_k$ when $i_k = 1$, and the right graph shows the proportion of $i_k = 1$.}
\label{fig:Trajectories}
\end{figure}

\begin{figure}
\center
\psfrag{x}[Bc][Bc][0.9]{$s$}
\psfrag{pidistr}[Bc][Bc][0.9]{$\mathbb{P}[i = 1\,|\,s,\vect\theta]$}
\psfrag{pi}[Bc][Bc][1.1]{$\vect\pi_{\vect\theta}^\mathrm{c}$}
\includegraphics[width=0.45\textwidth,clip]{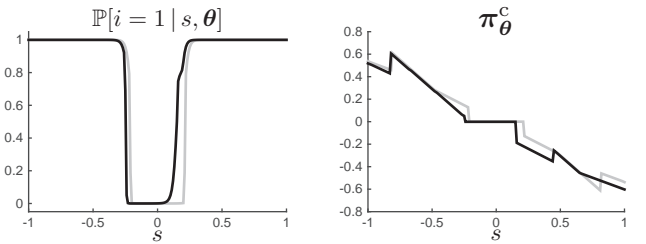}
\caption{Policy before (light grey) and after (black) the learning. The left graph shows the Softmax policy \eqref{eq:Softmax} as a function of the state $s$, giving the probability of selecting $i = 1$ for a given state $s$. The right graph shows the MPC policy (without the stochastic choice of integer variable).}
\label{fig:Policy}
\end{figure}

\begin{figure}
\center
\psfrag{RLstep}[Bc][Bc][.7]{RL step}
\psfrag{J}[Bc][Bc][1.]{$J(\pi_{\vect\theta})/J(\pi_{\vect\theta_0})$}
\includegraphics[width=0.3\textwidth,clip]{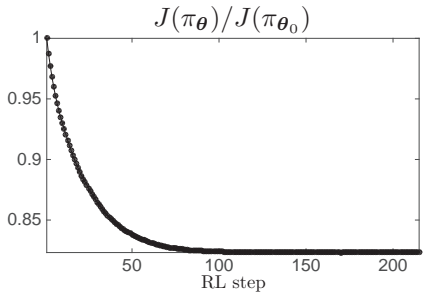}
\caption{Evolution of the closed-loop relative performance throughout the learning process. A reduction of the cost of over 15\% is achieved here from $\vect\theta_0$.}
\label{fig:Cost}
\end{figure}

The parameters $s_\mathrm{ref} = a^\mathrm{c}_\mathrm{ref} = 0$, $ w = 0.2$, $c=1$ were adopted for the baseline cost. The MPC scheme parameters were initialized using the same values, and using $b = 0$. Fig. \ref{fig:Trajectories} reports the trajectories of the system at the beginning and end of the learning process, showing how performance is gained by bringing the state trajectories in the interval $[-0.2,0.2]$. Fig. \ref{fig:Policy} reports the policy for the continuous and integer inputs, showing how RL reshapes the MPC policy for a better closed-loop performance. Fig. \ref{fig:Gradients} reports the estimated policy gradients via the compatible approximation \eqref{eq:MainResult} and directly via \eqref{eq:HybridPolicyGradient}, showing a match predicted by Prop. \ref{Prop:Main}. Fig. \ref{fig:Cost} reports the closed-loop performance of the MPC controller, calculated from  $J\left(\vect\pi_{\vect \theta}\right) = V_{\vect\pi_{\vect \theta}}(\vect s_0)$, and shows the performance gain obtained via the learning. Fig. \ref{fig:Parameters} shows the MPC parameter evolution through the learning process.

\newcommand{\FS}{0.5}
\newcommand{\Exp}{8}

\begin{figure}
\center
\psfrag{RLstep}[Bc][Bc][0.6]{RL step}
\psfrag{xref}[Bc][Bc][1.]{$\nabla_{x_\mathrm{ref}}J$}
\psfrag{uref}[Bc][Bc][1.]{$\nabla_{u_\mathrm{ref}}J$}
\psfrag{b}[Bc][Bc][1.]{$\nabla_{b}J$}
\psfrag{w}[Bc][Bc][1.]{$\nabla_{w}J$}
\psfrag{cc}[Bc][Bc][1.]{$\nabla_{c}J$}
\includegraphics[width=\FS\textwidth,clip]{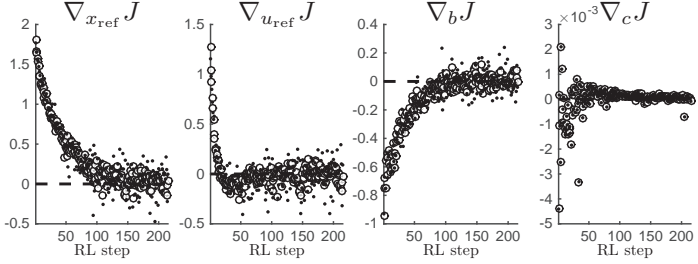}
\caption{Policy gradients throughout the learning process (iterations of the RL method). The dots display the policy gradient as obtained from \eqref{eq:MainResult}, while the circles display the policy gradients obtained from \eqref{eq:HybridPolicyGradient}.}
\label{fig:Gradients}
\end{figure}


\begin{figure}
\center
\psfrag{Refs}[Bc][Bc][1.]{$x_\mathrm{ref}$,$u_\mathrm{ref}$}
\psfrag{Int}[Bc][Bc][1.]{$w$}
\psfrag{Biases}[Bc][Bc][1.]{$b$}
\psfrag{Const}[Bc][Bc][1.]{$c$}
\psfrag{RLstep}[Bc][Bc][.7]{RL step}
\includegraphics[width=\FS\textwidth,clip]{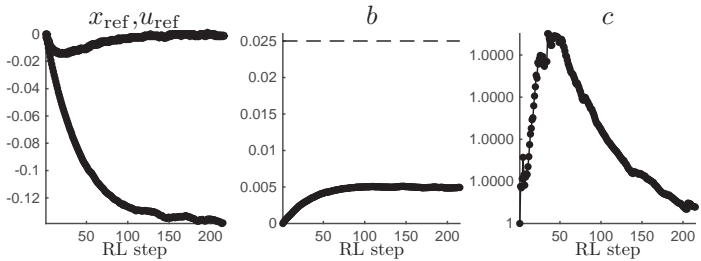}
\caption{Evolution of the MPC parameters throughout the learning process. The references are adjusted so that the system trajectories are better contained in the interval $[-0.2,0.2]$. The model bias $b$ does not match the value that a classic Prediction Error Method would deliver ($b = \mathbb E[n_k] = 0.025$, dashed line), while the cost associated to constraints is left unchanged.}
\label{fig:Parameters}
\end{figure}
\section{Discussion \& Conclusion} \label{sec:Conclusion}
This paper proposed an actor-critic approach to compute the policy gradient associated to policy approximations based on mixed-integer MPC schemes. The methodology is generic and applicable to linear, nonlinear and robust approaches. The paper proposes a hybrid stochastic-deterministic policy approach to generate the exploration and evaluate the policy gradient, avoiding the heavy computational expenses associated to using a stochastic policy approach on problems having continuous inputs and state constraints. A simple, compatible advantage function approximation is then proposed, tailored to our formulation and to MPC-based policy approximations. Some implementation details are provided, and the methods are illustrated on a simple example, providing a clear picture of how the proposed method is performing.

Future work will consider extensions to reduce the noise in the policy gradient estimation resulting from the choice of advantage function approximation, and will investigate techniques to integrate the stochastic policy and sensitivity computations with the branch-and-bound techniques used to solve the mixed-integer MPC problem. Future work will also investigate the potential of using the approaches detailed here to offer computationally less expensive approaches to solve the mixed-integer problem.
\bibliography{syscop}

\end{document}